\documentclass[conference]{IEEEtran}
\IEEEoverridecommandlockouts

\usepackage{cite}

\ifCLASSINFOpdf
\else
  \usepackage[dvips]{graphicx}
\fi
\usepackage[cmex10]{amsmath}
\usepackage{array}
\usepackage[tight,footnotesize]{subfigure}
\usepackage{url}
\usepackage{color}
\usepackage{amsthm}
\usepackage{amssymb}
\theoremstyle{plain}
\newtheorem{lemma}{\textbf{Lemma}}
\newtheorem{theorem}{Theorem}

\theoremstyle{definition}
\newtheorem{remark}{Remark}

\begin{document}
\title{Delay Minimization in Real-time Communications with Joint Buffering and Coding}

\author{
\IEEEauthorblockN{
Jesper H. S\o rensen, Petar Popovski and Jan \O stergaard
}
\IEEEauthorblockA{
Aalborg University, Department of Electronic Systems,
Fredrik Bajers Vej 7, 9220 Aalborg, Denmark
\\
E-mail: \{jhs,petarp,jo\}@es.aau.dk
\thanks{The research presented in this paper was supported by the Danish Council for Independent Research (Det Frie Forskningsr{\aa}d) DFF - $4184-00227B$}
}
}
\maketitle

\begin{abstract}
We present a closed-form expression for the minimal delay that is achievable in a setting that combines a buffer and an erasure code, used to mitigate the packet delay variance. The erasure code is modeled according to the recent information-theoretic results on finite block length codes. Evaluations reveal that accurate knowledge of the network parameters is essential for optimal operation. Moreover, it is shown that, when the network packet delay variance is large, the buffer delay becomes negligible. Therefore, in this case the delay budget should be spent mainly on the erasure code.
\end{abstract}
\IEEEpeerreviewmaketitle

\section{Introduction} \label{sec:introduction}
Low latency is set as a major goal in the next generation mobile communication systems. This is a result of an increased interest in interactive applications, such as voice over IP, video conversations, telemedicine, gaming, automated transport systems and virtual reality. The term \textit{tactile Internet} has been used for a level of connectivity that makes remote locations feel at a fingertip, leading to entirely new applications and industries. This vision can only be realized through giant leaps in terms of latency reduction in communication technologies.

Several recent works deal with communication with delay constraints \cite{nikaein2011,yilmaz2015}. Streaming is an example of an application that has received great attention \cite{chou2006,nafaa2008}. In such an application, data has a deadline after which the data is useless. Variance in the data packet delay leads to stochastic late arrivals, which is a major challenge. As late arrival is equivalent to not arriving at all, the deadline enforcement can thus be considered as a creation of an \emph{erasure channel}. This insight led to several works on erasure coding for streaming applications \cite{leong2013,martinian2007,badr2013}.

Classical information theory deals with the relation between the achievable coding rate and the channel parameter, e.g. the erasure probability in an erasure channel. A fundamental requirement is that arbitrarily low probability of decoding error can be achieved, while allowing the block length to grow towards infinity. However, in practical systems, arbitrarily low probability of error is not necessary, while a codeword length, i.e. delay, that goes towards infinity is intolerable. A different information-theoretic approach was developed in \cite{polyanskiy2010}, where the coding rate is expressed as a function of the channel parameter at a desired blocklength and decoding error probability. This serves to quantify the loss of capacity, when considering practical codes of finite blocklength. Similarly, this theory tells us the necessary blocklength, which is proportional to the delay, when applying a particular coding rate and requiring a certain decoding error probability. 

In this work we apply the finite blocklength theory in a joint design of a buffer and an erasure code to minimize the delay while providing the desired reliability. The idea of coupling the design of a buffer and forward error correction was first proposed in \cite{rosenberg2000} and later extended in \cite{boutremans2003}. These works proposed adaptive algorithms involving simple repetition codes, where redundant versions of audio packets are piggybacked on subsequent packets. Our work applies information-theoretic results that hold for arbitrary erasure codes.

\section{System Model and Problem Definition}\label{sec:sysmodel}
We consider a real-time communication system with a single source and a single receiver. Messages are generated at the source with a fixed time interval $t_s\in\mathbb{R}^+$, such that message $j$ is generated at time $t_j=jt_s$. Each message must be decoded no later than $t_d\in\mathbb{R}^+$ from the generation time, whereby the deadline of message $j$ is $jt_s+t_d$. All messages have the same size and can be transmitted as independent data packets e.g. over the Internet. These transmissions experience a delay, which is the sum of a number of contributing factors, e.g. propagation, queuing etc. 

\begin{figure}[t]
 \centering
 \includegraphics[width=1\columnwidth]{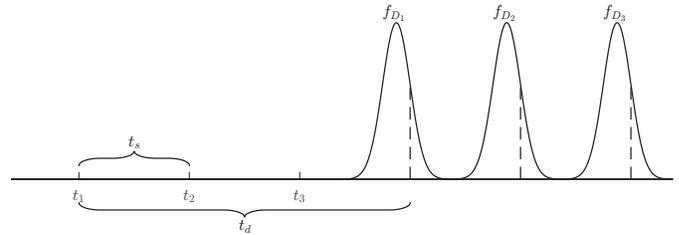}
 \vspace{-0.7cm}
 \caption{The relative timing of the messages and their deadlines.}
 \label{fig:system}
\end{figure}

We model the delay of packet $j$ as a random variable $D_j$. As in the prior work \cite{rosenberg2000,boutremans2003}, we assume that the packet delays are independent, i.e. $D_j$ are i.i.d. variables with probability density function, $f_{D_j}$, mean value $\mu$ and higher moments $\mathbb{E}[(D_j-\mu)^r]$ for $r>1$ that are known at the transmitter.\footnote{In this work, $\mathbb{E}[(D_j-\mu)^r]$ for $r>1$ represent a sufficient knowledge at the transmitter. Knowing only them, but not $\mu$, is a weak assumption, since it only requires estimations of relative delays, which is much easier in practice compared to absolute delay estimation, which is required for knowing $\mu$.} The $j-$th packet is received at time $jt_s+D_j$, see Fig.~\ref{fig:system} for an illustration. If a packet is received after the deadline, i.e. $D_j>t_d$, it is considered as an erasure, whose probability we denote $\epsilon$. The message deadline, $t_d$, is not considered fixed and dictated by the application, but rather a design parameter, which can be tuned in order to achieve a desired quality of service. In this regard, it is assumed that the receiver (but not the transmitter) knows the expected arrival time of the $j$'th packet $\mathbb{E}[jt_s+D_j]=jt_s+\mu$, such that an arbitrary deadline can be enforced. We introduce the relative deadline $t_{\Delta}=t_d-\mu$.

\remark{The considered system model is very appropriate for streaming applications, such as Voice over IP (VoIP) or interactive video communication. It also finds application in real-time networked control systems, e.g. smart grid systems. The emerging Internet of things is expected to bring a plethora of applications, for which this system model is relevant.}

Depending on the application, messages received prior to the deadline will either be applied immediately or stored in a buffer until the exact deadline. As an example, VoIP systems must provide a smooth delivery of the audio, such that data must be applied with fixed intervals, i.e. at the deadlines. Higher moments of the delay is often termed \emph{jitter} in such systems, and the remedy is accordingly termed a \emph{jitter buffer}. In this work, we adopt the notion that messages received prior to the deadline is handled by a buffer, such that the $j-$th message spends the time $t_d-D_j$ in the buffer. This implies that all messages exiting the buffer are delayed by $t_d$ from their generation time $jt_s$. Naturally, the size of the buffer must be proportional to $t_{\Delta}$, however, memory is assumed to be available in abundance. We are solely concerned with the delay implication of increasing $t_{\Delta}$.

The problem we consider is how to minimize the delay introduced by a buffer and a rate $R$ erasure code, given certain packet delay variance (PDV), expressed through the higher moments of $D_j$, and a certain requirement on the decoding error probability of the erasure code.
\section{Joint Buffering and Coding}\label{sec:scheme}
The dimensioning of the buffer is equivalent to a choice of $t_d$, since any packets arriving before the deadline is stored in a buffer until its deadline. For a particular $t_d$, the resulting erasure probability is $\epsilon=1-F_D(t_d)$, where $F_D$ is the cumulative distribution function (CDF) of the network delay.

We further apply a packet level erasure code with rate $R=\frac{k}{n}$, where $n$ is the number of packets forming a codeword, which is generated as a function of $k$ original packets, each representing a message. The code is a linear block code in a sufficiently large field, such that it is a reasonable assumption that any subset of $k$ out of the $n$ packets is sufficient for successful decoding. We define $\delta$ as the probability of not being able to decode a codeword, i.e. more than $n-k$ erasures occurred. We refer to this as the decoder error probability.

\remark{When encoding the $k$ messages into $n$ packets, it is not immediately clear what the deadline of these packets should be. They are linear combinations of the $k$ messages, each having their own deadline. Since a decoder error probability of at most $\delta$ is tolerated, we must guarantee this error probability for each message. Hence, the deadline of encoded packets must be defined by the earliest of the $k$ messages. We therefore enforce the deadline $\ell t_s+t_d$ on all $n$ packets, where $\ell t_s$ is the generation time of the first of the $k$ messages. Hence, a delay of $kt_s$ is introduced by the erasure code.\footnote{$kt_s$ is the maximum delay experienced only by the first of the $k$ packets. In fact, the last of the $k$ packets experience no added delay.}$^,$\footnote{The erasure code also adds a delay proportional to $n-k$, the number of redundant packets added by the code, due to the transmission of more packets. This is considered negligible among the other delay components.}}

The relation between $k$, $\delta$, $R$ and $\epsilon$ was derived in \cite{polyanskiy2010} and allows us to quantify the delay introduced by an erasure code with rate $R$, that achieves the desired $\delta$ in an erasure channel with erasure probability $\epsilon$. Based on a Gaussian approximation \cite{polyanskiy2010} of the number of erasures in a codeword, the relation is
\begin{align}\label{k1}
k &= n(1-\epsilon)-\sqrt{n\epsilon(1-\epsilon)}Q^{-1}(\delta) + \mathcal{O}(1),
\end{align}
where $Q$ is the complementary Gaussian CDF. When disregarding the higher order terms, \eqref{k1} can be rewritten as
\begin{align}\label{eq:k}
k &= R\left(\frac{\sqrt{\epsilon(1-\epsilon)}Q^{-1}(\delta)}{1-\epsilon-R}\right)^2.
\end{align}
%
%
The total delay of the system is $t_d+kt_s=\mu+t_{\Delta}+kt_s$. Since $\mu$ is a constant given by the network, minimizing the total delay is equivalent to minimizing $t_{\Delta}+kt_s$. By normalizing with the message interval, we can define the system delay as 
\begin{align}\label{eq:gamma}
\gamma&=\frac{t_{\Delta}}{t_s}+k.
\end{align}
The objective in this work is to minimize $\gamma$ for particular choices of $\delta$ and $R$. This objective brings an interesting tradeoff: \textsl{Given a reliability target $\delta$, how should the burden be divided between the buffer and the erasure code?} The increase in the buffer leads to increase in the deadline $t_d$, which results in a lower erasure probability. The latter implies that the delay, i.e. the blocks size $k$ of the erasure code is decreased, while \eqref{k1} implies that lower block size $k$ leads to less efficient erasure code. Hence, minimal $\gamma$ arises by optimally ``investing'' in deadline $t_d$ and block size $k$ that offer a certain erasure correction capability. The optimal balance of the two methods for combating PDV in the case of $D_j$ following the Gaussian distribution is analyzed in the remainder of this section.


\subsection{Gaussian Assumption}\label{gaussian}
When assuming $D_j$ follows a Gaussian distribution with mean $\mu$ and standard deviation $\sigma$, we can express the erasure probability using the $Q$-function:

\begin{align}\label{eq:epsilon}
\epsilon &= Q\left(\frac{t_{\Delta}}{t_s\sigma}\right).
\end{align}
Combining \eqref{eq:k}, \eqref{eq:gamma} and \eqref{eq:epsilon}, we get the following expression of the total system delay:
\begin{align}\label{exactgamma}
\gamma &= \sigma Q^{-1}(\epsilon)+R\left(\frac{\sqrt{\epsilon(1-\epsilon)}Q^{-1}(\delta)}{1-\epsilon-R}\right)^2.
\end{align}
We wish to minimize $\gamma$ through an optimal combination of the receiver buffer and the erasure code. A combination is represented by $\epsilon$, which dictates the size of the receiver buffer through \eqref{eq:epsilon} and the block length of the erasure code through \eqref{eq:k}. Minimizing $\gamma$ with respect to $\epsilon$ is infeasible analytically due to $Q^{-1}(\epsilon)$, which has no closed form expression. We therefore apply the approximation \cite[pp. 331--336]{betten2008}:
\begin{align}\label{approx}
Q^{-1}(\epsilon) &\approx \frac{\sqrt{2}\tanh^{-1}(1-2\epsilon)}{u},
\end{align}
for suitably chosen $u$, which balances accuracy and range of validity. We note that \eqref{exactgamma} is itself an approximation, as $\mathcal{O}(1)$ from \eqref{k1} is unknown but considered negligible.
We can then define the approximated system delay, $\tilde{\gamma}(\epsilon)$:
\begin{align}
\tilde{\gamma}(\epsilon) &:= \frac{\sqrt{2}\sigma \tanh^{-1}(1-2\epsilon)}{u}+R\left(\frac{\sqrt{\epsilon(1-\epsilon)}Q^{-1}(\delta)}{1-\epsilon-R}\right)^2.
\end{align}
This enables us to state the optimization problem:
\begin{align}\label{eq:min}
\underset{\epsilon}{\text{minimize}} & & \tilde{\gamma}(\epsilon), \notag \\
\text{subject to} & & 0 \le \epsilon \le 1-R.
\end{align}
\begin{lemma} \label{lemma:onemin} The objective function in the optimization problem in \eqref{eq:min}, $\tilde{\gamma}(\epsilon)$, has at least one local minimum in the interval $0\le\epsilon\le 1-R$ for $0<R\le 0.5$, $\sigma>0$, $\delta>0$.
\end{lemma}
\begin{IEEEproof}
We show that $\tilde{\gamma}'$ satisfies $\tilde{\gamma}'(0)<0$ and $\tilde{\gamma}'(1-R)>0$, where $\tilde{\gamma}'(\epsilon)$ is the derivative of $\tilde{\gamma}(\epsilon)$. This implies that $\tilde{\gamma}$ has at least one minimum in the desired interval, $0\le\epsilon\le 1-R$. The derivative is found as
\begin{align}\label{eq:derivative}
\tilde{\gamma}'(\epsilon)&=\frac{\sigma}{\sqrt{2} u (\epsilon-1) \epsilon}-\frac{Q^{-1}(\delta)^2 R (R (2 \epsilon-1)-\epsilon+1)}{(R+\epsilon-1)^3}.
\end{align}
It is seen from the first fraction in \eqref{eq:derivative} that $\tilde{\gamma}'(0)=-\infty$ and from the second fraction that $\tilde{\gamma}'(1-R)=\infty$.
\end{IEEEproof}
\begin{lemma} \label{lemma:convex} The optimization problem in \eqref{eq:min} is convex for $0<R\le 0.5$, $\sigma>0$, $\delta>0$ and $0\le\epsilon\le 1-R$.
\end{lemma}
\begin{IEEEproof}
$\tilde{\gamma}(\epsilon)$ is convex for $0\le\epsilon\le 1-R$ if its second derivative, $\tilde{\gamma}^{''}(\epsilon)$, is positive in that interval under the constraints on $R$, $\sigma$ and $\delta$. The second derivative is found as
\begin{align}
\tilde{\gamma}''(\epsilon) &= \frac{2\left(Q^{-1}(\delta)\right)^2R\left((2R-1)\epsilon-R^2+1\right)}{(R+\epsilon-1)^4} \notag \\
                              &\quad +\frac{\sigma(1-2\epsilon)}{\sqrt{2}u(\epsilon-1)^2\epsilon^2},
\end{align}
which is positive, since all factors in the two fractions are positive for $0<R \le 0.5$, $\sigma>0$, $\delta>0$ and $0\le\epsilon\le 1-R$.
\end{IEEEproof}
Lemma~\ref{lemma:onemin} states that a local minimum exists in the interval of interest and from the convexity stated in Lemma~\ref{lemma:convex} we have that such a local minimum is guaranteed to be a global minimum. We can therefore state the following theorem:
\begin{theorem} \label{theorem:opteps} The optimal erasure probability, $\epsilon^{\star}$, leading to minimal latency, $\tilde{\gamma}^{\star}$, for given $\delta$, $u$, $\sigma$, and $R$ is the unique solution in the interval $0\leq \epsilon\leq 1-R$ to the following:
\begin{align}\label{eq:poly3}
0&=a\epsilon^3+b\epsilon^2+c\epsilon+d,\notag \\
a&=-2RQ^{-1}(\delta)^2u(2R-1)+\sqrt{2}\sigma,\notag \\
b&=2RQ^{-1}(\delta)^2u(2R-1)-2RQ^{-1}(\delta)^2u(1-R)\notag \\
&\quad+\sqrt{2}\sigma(3R-3),\notag \\
c&=2RQ^{-1}(\delta)^2u(1-R)\notag \\
&\quad+\sqrt{2}\sigma\left((R-1)(2R-2)+(R-1)^2\right),\notag \\
d&=\sqrt{2}\sigma(R-1)^3.
\end{align}
\end{theorem}
\begin{IEEEproof}
The cubic equation in \eqref{eq:poly3} is reached by rearranging the right hand side of \eqref{eq:derivative} and equating to zero. The solution to the optimization problem in \eqref{eq:min} can therefore be found using the cubic formula for the roots of a polynomial of third degree with the coefficients in \eqref{eq:poly3}. The root of interest is the single one lying in the interval $0\le\epsilon\le 1-R$.
\end{IEEEproof}

\section{Numerical Results} \label{sec:results}
\begin{figure}[t]
 \centering
 \includegraphics[width=1\columnwidth]{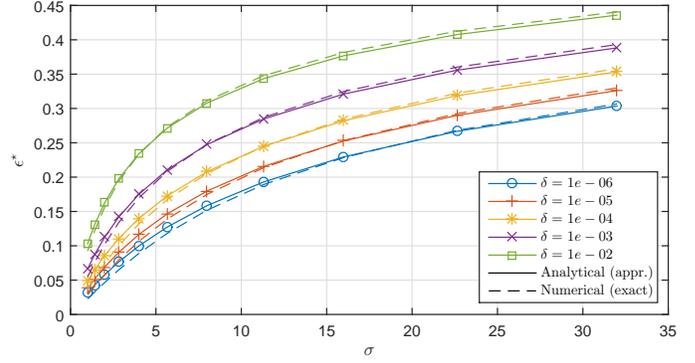}
\vspace{-0.7cm}
 \caption{Optimal erasure probability as a function of the standard deviation of the network delay for $R=\frac{1}{3}$.}
 \label{fig:eps_sig}
\end{figure}

In Fig.~\ref{fig:eps_sig}, $\epsilon^{\star}$ is plotted as a function of $\sigma$, when $R=\frac{1}{3}$. The parameter $u=1.2028$ is chosen as suggested in \cite[pp. 331--336]{betten2008}. The analytical results achieved by solving \eqref{eq:min} using \eqref{eq:poly3} are plotted with solid lines. Moreover, numerical results achieved using the Nelder-Mead simplex algorithm \cite{nelder1965} on the exact expression in \eqref{exactgamma} are included using dashed lines. The curves are seen to match well, which validates the approximation in \eqref{approx}. Moreover, the figure shows that $\epsilon^{\star}$ is quite sensitive to $\sigma$, which indicates that accurate knowledge of the network conditions is important for this optimization.

\begin{figure}[t]
 \centering
 \includegraphics[width=1\columnwidth]{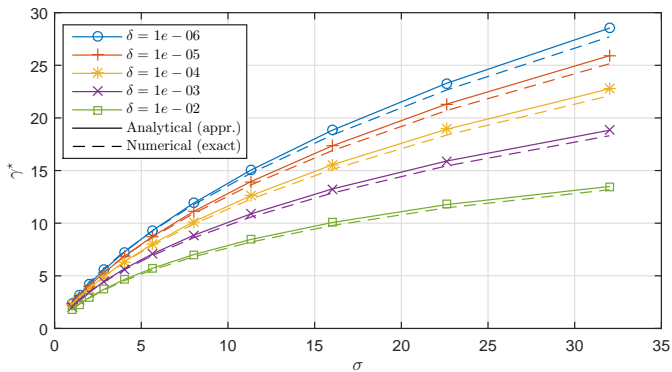}
\vspace{-0.7cm}
 \caption{The minimum system delay as a function of the standard deviation of the network delay for $R=\frac{1}{3}$.}
 \label{fig:lat_sig}
\end{figure}

The resulting $\gamma^{\star}$ is plotted in Fig.~\ref{fig:lat_sig}. Again, both analytical and numerical results are included and shown to match well. These results show that an increased PDV implies an increased system delay, which follows intuition.

\begin{figure}[t]
 \centering
 \includegraphics[width=1\columnwidth]{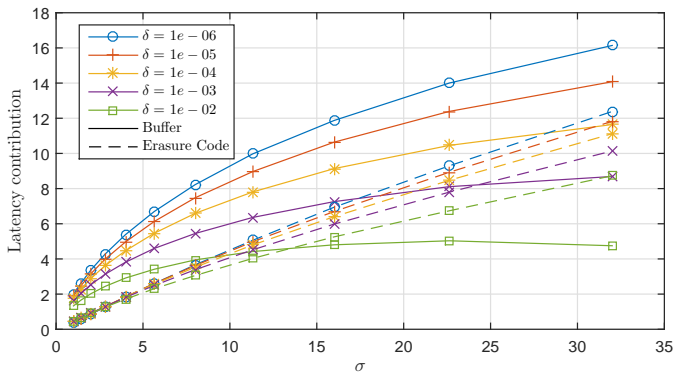}
\vspace{-0.7cm}
 \caption{Delay contributions from the buffer and the erasure code.}
 \label{fig:contributions}
\end{figure}

Fig.~\ref{fig:contributions} shows the minimum delays from the numerical results in Fig.~\ref{fig:lat_sig} split into two contributions, one from the buffer and one from the erasure code. It shows that the buffer contribution has a decaying increase as $\sigma$ increases. At high $\sigma$ it even starts to decrease. This tells that the optimal approach at high $\sigma$ is to put most of the burden on the erasure code.

\begin{figure}[t]
 \centering
 \includegraphics[width=1\columnwidth]{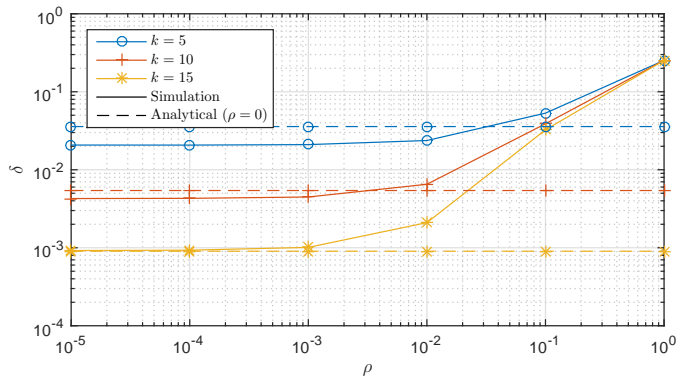}
\vspace{-0.7cm}
 \caption{Comparison of simulation results and analytical results for different values of $k$.}
 \label{fig:sim}
\end{figure}

Finally, simulation results are shown for a system with buffer delay $\frac{t_{\Delta}}{t_s}=8$ and erasure codes with $R=\frac{1}{2}$ and $k=\{5,10,15\}$, such that $n=\{10,20,30\}$. The code is a random linear code with field size $2^8$, which is practical in an implementation, since coding coefficients can be stored as bytes. Packet delays are modeled as correlated Gaussian random variables, hence $D_j \sim\ \mathcal{N}(\boldsymbol\mu,\, \boldsymbol\Sigma)$, where $\boldsymbol\mu_i=\mu$ for $i=1\dots n$ and $\boldsymbol\Sigma_{ij}=\sigma^2$ for $i=j$ and $\boldsymbol\Sigma_{ij}=\rho\sigma^2$ for $i\neq j$, $i=1 \dots n$, $j=1\dots n$ and $0 \leq \rho \leq 1$. Without loss of generality, we consider the case of $\mu=0$, since any non-zero mean value will merely add a constant delay to the system, which does not influence our scheme. We moreover consider the case of $\sigma=12$. Fig.~\ref{fig:sim} shows $\delta$ as a function of $\rho$, the correlation coefficient. The figure also shows analytical results, which can be found by combining equations \eqref{eq:k} and \eqref{eq:epsilon} and isolating $\delta$. Note that the analysis has been performed for the special case of $\rho=0$. Comparing simulations and analysis, it is evident that the analysis is increasingly valid as $k$ increases. The figure also shows that the analysis is valid at low correlation coefficients, but breaks down when the correlation becomes significant, which is expected.   

\section{Conclusions and Future Work}\label{sec:conclusions}
We have analyzed the problem of minimizing the delay in a joint buffering and erasure correction scheme for mitigation of packet delay variance. Numerical results, for the case of Gaussian distributed packet delay, show that the optimal operating point is highly sensitive to the variance of the packet delay. This underlines the importance of a joint optimization of the buffer and erasure code. Moreover, results show that under bad channel conditions, i.e. high packet delay variance, it is more advantageous to spend delay on the erasure code in order to achieve the target reliability. Future work is oriented around more realistic packet delay models, both considering heavy tail distributions and correlation between packets.

\bibliographystyle{ieeetr}
\bibliography{bibliography}

\begin{thebibliography}{10}

\bibitem{nikaein2011}
N.~Nikaein and S.~Krea, ``Latency for real-time machine-to-machine
  communication in lte-based system architecture,'' in {\em Wireless Conference
  2011 - Sustainable Wireless Technologies (European Wireless), 11th European},
  pp.~1--6, April 2011.

\bibitem{yilmaz2015}
O.~N.~C. Yilmaz, Y.~P.~E. Wang, N.~A. Johansson, N.~Brahmi, S.~A. Ashraf, and
  J.~Sachs, ``Analysis of ultra-reliable and low-latency 5g communication for a
  factory automation use case,'' in {\em 2015 IEEE International Conference on
  Communication Workshop (ICCW)}, pp.~1190--1195, June 2015.

\bibitem{chou2006}
P.~A. Chou and Z.~Miao, ``Rate-distortion optimized streaming of packetized
  media,'' {\em IEEE Transactions on Multimedia}, vol.~8, pp.~390--404, April
  2006.

\bibitem{nafaa2008}
A.~Nafaa, T.~Taleb, and L.~Murphy, ``Forward error correction strategies for
  media streaming over wireless networks,'' {\em IEEE Communications Magazine},
  vol.~46, pp.~72--79, January 2008.

\bibitem{leong2013}
D.~Leong, A.~Qureshi, and T.~Ho, ``On coding for real-time streaming under
  packet erasures,'' in {\em Information Theory Proceedings (ISIT), 2013 IEEE
  International Symposium on}, pp.~1012--1016, July 2013.

\bibitem{martinian2007}
E.~Martinian and M.~Trott, ``Delay-optimal burst erasure code construction,''
  in {\em 2007 IEEE International Symposium on Information Theory},
  pp.~1006--1010, June 2007.

\bibitem{badr2013}
A.~Badr, A.~Khisti, W.~T. Tan, and J.~Apostolopoulos, ``Streaming codes for
  channels with burst and isolated erasures,'' in {\em INFOCOM, 2013
  Proceedings IEEE}, pp.~2850--2858, April 2013.

\bibitem{polyanskiy2010}
Y.~Polyanskiy, H.~V. Poor, and S.~Verdu, ``Channel coding rate in the finite
  blocklength regime,'' {\em IEEE Transactions on Information Theory}, vol.~56,
  pp.~2307--2359, May 2010.

\bibitem{rosenberg2000}
J.~Rosenberg, L.~Qiu, and H.~Schulzrinne, ``Integrating packet fec into
  adaptive voice playout buffer algorithms on the internet,'' in {\em INFOCOM
  2000. Nineteenth Annual Joint Conference of the IEEE Computer and
  Communications Societies. Proceedings. IEEE}, vol.~3, pp.~1705--1714 vol.3,
  Mar 2000.

\bibitem{boutremans2003}
C.~Boutremans and J.~Y.~L. Boudec, ``Adaptive joint playout buffer and fec
  adjustment for internet telephony,'' in {\em INFOCOM 2003. Twenty-Second
  Annual Joint Conference of the IEEE Computer and Communications. IEEE
  Societies}, vol.~1, pp.~652--662 vol.1, March 2003.

\bibitem{betten2008}
J.~Betten, {\em Creep Mechanics}.
\newblock Springer, 2008.

\bibitem{nelder1965}
J.~A. Nelder and R.~Mead, ``A simplex method for function minimization,'' {\em
  The Computer Journal}, vol.~7, no.~4, pp.~308--313, 1965.

\end{thebibliography}

\end{document}